\documentclass[aps,twocolumn,pra,superscriptaddress,showpacs,showkeys,floatfix]{revtex4-1}

\usepackage{amsmath,amsfonts,amssymb,graphicx}
\usepackage{hyperref}
\usepackage{hyperref}
\usepackage{color}
\usepackage{subfigure}
\usepackage{epsfig}
\usepackage{subfigure}
\usepackage[normalem]{ulem} 
\usepackage{braket}

\begin{document}

\title{Amending entanglement-breaking channels via intermediate unitary operations}

\author{\'A. Cuevas}
\affiliation{Universit\`a di Roma La Sapienza, Dipartimento di Fisica, 00185, Rome, Italy}

\author{A. De Pasquale}
\affiliation{NEST, Scuola Normale Superiore and Istituto Nanoscienze-CNR, 56127 Pisa, Italy}

\author{A. Mari}
\affiliation{NEST, Scuola Normale Superiore and Istituto Nanoscienze-CNR, 56127 Pisa, Italy}

\author{A. Orieux}
\affiliation{Universit\`a di Roma La Sapienza, Dipartimento di Fisica, 00185, Rome, Italy}
\affiliation{LIP6, CNRS, Universit\'e Pierre et Marie Curie, Sorbonne Universit\'es, 75005 Paris, France}
\affiliation{IRIF, CNRS, Universit\'e Paris Diderot, Sorbonne Paris Cit\'e, 75013 Paris, France}

\author{S. Duranti}
\affiliation{Universit\`a di Roma La Sapienza, Dipartimento di Fisica, 00185, Rome, Italy}
\affiliation{Universti\`a degli Studi di Perugia, Dipartimento di Fisica e Geologia, 06123, Perugia, Italy}

\author{M. Massaro}
\affiliation{Universit\`a di Roma La Sapienza, Dipartimento di Fisica, 00185, Rome, Italy}
\affiliation{University of Paderborn, Department of Physics, 33098 Paderborn, Germany}

\author{A. Di Carli}
\affiliation{Universit\`a di Roma La Sapienza, Dipartimento di Fisica, 00185, Rome, Italy}
\affiliation{University of Strathclyde, Department of Physics, Scottish Universities Physics Alliance (SUPA), Glasgow G4 0NG, United Kingdom}

\author{E. Roccia}
\affiliation{Universit\`a di Roma La Sapienza, Dipartimento di Fisica, 00185, Rome, Italy}
\affiliation{Universit\`a degli Studi Roma Tre, Dipartimento di Fisica, 00146, Rome, Italy}

\author{J. Ferraz}
\affiliation{Universit\`a di Roma La Sapienza, Dipartimento di Fisica, 00185, Rome, Italy}
\affiliation{Universidade Federal Rural de Pernambuco, Departamento de Física, 52171-900, Recife, Brazil}

\author{F. Sciarrino}
\affiliation{Universit\`a di Roma La Sapienza, Dipartimento di Fisica, 00185, Rome, Italy}

\author{P. Mataloni}
\affiliation{Universit\`a di Roma La Sapienza, Dipartimento di Fisica, 00185, Rome, Italy}

\author{V. Giovannetti}
\affiliation{NEST, Scuola Normale Superiore and Istituto Nanoscienze-CNR, 56127 Pisa, Italy}

\begin{abstract}
We report a bulk optics experiment demonstrating the possibility of restoring the entanglement distribution through noisy quantum channels by inserting  a suitable unitary operation ({\it filter}) in the middle of the transmission process. 
We focus on two relevant classes of single-qubit channels  consisting in repeated applications of {\it rotated phase damping} or {\it rotated amplitude damping} maps, both modeling the combined Hamiltonian and dissipative dynamics of the polarization state of single photons. Our results show that interposing a unitary filter between two  noisy channels can significantly improve entanglement transmission. This proof-of-principle demonstration could be generalized to many other physical scenarios where entanglement-breaking communication lines may be amended by unitary filters.  \end{abstract}

\maketitle

\section{Introduction}

Real quantum communication channels are typically not perfect transmission lines, since they usually introduce different kinds of noises given by the intrinsic mechanisms that transfer information or by external perturbations. Physically, such channels disturb the transmitted messages by gradually degrading the information  along their structures \cite{barnum,schumacher,nielsen97}, and this effect may be particularly severe when entangled qubits are propagated through them. The extreme limit is represented by {\it entanglement-breaking} (EB) channels \cite{horodecki2003} which are so noisy to be useless for entanglement distribution even exploiting distillation techniques \cite{bennet1996}, with direct consequences on the associated classical or quantum capacities \cite{channelcapacity}. 

In many practical situations, quantum channels can be represented as the consecutive application of a given elementary map $\Phi$ repeated $n$ times, where $n$ is a positive integer. In this cases, the full channel is given by
\begin{align}
\Phi^n=\underbrace{\Phi\circ \Phi \circ \cdots \circ \Phi}_{n \;\; \mbox{{\small times}} }
\end{align}	
and the integer $n$ can be interpreted as the effective length of the transmission line. For example if the elementary map $\Phi$ corresponds to the spatial propagation of a quantum state along a physical medium of length $l$, then the total length of the channel is $n\, l$. Alternatively if $\Phi$ models the dissipative evolution of the system lasting  an elementary time interval $\tau$, then $n\, \tau$ represents the total time duration of the whole process. In these cases one may ask what is the maximum entanglement propagation length or, equivalently, what is the minimum $n=k$ such that $\Phi^k$ is EB. Such a number $k$ can be seen as a sort of noise quantifier for the elementary map $\Phi$  and corresponds to  its  {\it entanglement-breaking order}  originally defined in \cite{depasquale2012}, and further investigated in \cite{depasquale2013,lami2015,cutandpaste}.

In this work we give an experimental proof-of-principle demonstration that the propagation length can be increased by placing an intermediate unitary operation (filter) between the two elementary maps $\Phi$. More precisely, we implement two different examples (built up exploiting single-qubit phase-damping and amplitude-damping channels respectively) of an elementary map $\Phi$ such that:
\begin{align}
&\Phi\circ \Phi  \; {\rm is\ EB,} \label{filt1}\\
&\Phi \circ \mathcal F  \circ \Phi  \; {\rm is\ not\ EB,} \label{filt2}
\end{align}
for a suitable unitary operation $\mathcal F$. In practice, the action of the filter $\mathcal F$ is to ``amend" the otherwise EB communication line, by properly acting in the middle of the transmission process. Therefore the length of the channel up to which quantum correlations are preserved is increased. This idea has been theoretically introduced in~\cite{depasquale2012}.
Together with the recent results reported in Ref.\ \cite{cutandpaste} in which dissipative correcting operations have been considered, the present work represents the first experimental demonstration of such an entanglement recovery technique.

The manuscript is organized as follows. In Section \ I, we present the idea and the theoretical model of the two experiments: the first one based on rotated phase damping maps and the second one based on rotated amplitude damping maps. In Section II we present all the details of the experimental implementation. In Section III we report and discuss the experimental results. Eventually, in Section V we draw some conclusions. 

\section{Theoretical model of the experiment}

In this section we present the two experimental schemes that we used to verify the unitary filtering phenomenon compactly summarized  by Eq.s\ \eqref{filt1} and \eqref{filt2}. For both experiments we focused on single-qubit channels acting on the polarization state of individual photons. The vertical and horizontal polarizations of the photon form a basis $\{ | V \rangle, | H \rangle \}$  of a two-dimensional Hilbert space. In this basis, a generic quantum state can be expressed as a $2\times2$ density matrix $\rho \in \mathbb C^2$, such that $\rho\ge 0$ and $Tr[\rho]=1$.

As real candidates for the elementary map $\Phi$ appearing in Eq.s\ \eqref{filt1}, we considered the {\it rotated phase damping} and the {\it rotated amplitude damping} maps, respectively given by the following composition of operations:
\begin{align}
\Phi_{PD}=\Lambda_{\pi/8}\circ\Gamma, \\
\Phi_{AD}=\Lambda_{\pi/4}\circ\Sigma ,
\end{align}
where $\Lambda_{\pi/8}$ and  $\Lambda_{\pi/4}$ are unitary rotations defined by:
\begin{equation} \label{rotation}
\Lambda_\theta(\rho)= R_\theta \rho R_\theta,   \quad R_\theta=\left(\begin{matrix}
\cos(2\theta) & -\sin(2\theta)\\
-\sin(2\theta) & -\cos(2\theta)\end{matrix} \right),
\end{equation}
$\Gamma$ is the phase-damping channel  \cite{nielsen-chuang,desurvire} with damping parameter $p\in[0,1]$:
\begin{equation} \label{PDmap}
\Gamma(\rho)=\left(1-\frac{p}{2}\right)\rho+\frac{p}{2}\sigma_{z} \rho\sigma_{z}, \quad \sigma_z=R_{\theta=0},
\end{equation}
while $\Sigma$ is the amplitude damping channel  \cite{nielsen-chuang,desurvire} with parameter $\eta\in[0,1]$:
\begin{align}\label{ADmap}
\Sigma(\rho)=&E_1
\rho
E_1^\dagger
+E_2
\rho
E_2^\dagger\,,
\end{align}
with $E_1= \left(\begin{matrix}
1 & 0\\
0 & \sqrt{1-\eta}\end{matrix}\right)$ and $E_2=\left(\begin{matrix}
0 & \sqrt{\eta}\\
0 & 0\end{matrix}\right)$.
In both cases ($\Phi=\Phi_{PD}$ and $\Phi=\Phi_{AD}$), as a potential candidate for the generic unitary filter appearing in \eqref{filt2},  we use 
\begin{equation}
\mathcal F= \Lambda_{\varphi} ,
\end{equation}
where $\varphi$ is an angle that we are going to optimize and $\Lambda_\varphi$ is a unitary operation defined analogously to $\Lambda_\theta$ as in Eq.\ \eqref{rotation}.
Our aim is to experimentally verify that there exists values/intervals of the damping parameters $p$, $\eta$ and of the filter angle $\varphi$ such that both
conditions \eqref{filt1} and \eqref{filt2} are fulfilled, demonstrating that the unitary filter succeeds in increasing the entanglement propagation distance.

Before presenting the details of the experimental implementation, we anticipate here  how the previous three maps \eqref{rotation} \eqref{PDmap} and \eqref{ADmap} can be realized using standard bulk optics elements such as beam splitters, phase plates {\it etc.}, and how one can experimentally test whether a given sequence of maps is entanglement breaking or not.
The implementation of the unitary rotation $\Lambda_\varphi$ is very simple since it corresponds to the application of a $\lambda/2$ phase plate rotated by an angle $\varphi$ around the propagation axis of the photon. The phase-damping channel \eqref{PDmap} can also be implemented quite easily as a probabilistic switch between the identity operation $\mathbb{I}$ and the map $\sigma_{z}$. The application of the amplitude-damping channel \eqref{ADmap} to a polarization qubit is instead less straightforward but can still be simulated with a suitable interferometric scheme as explained in the next section.  Finally 
we recall that, in order to test whether a channel is entanglement breaking or not, it is sufficient to apply it to a subsystem of a maximally entangled state and check the separability of the output state \cite{horodecki2003}: 
\begin{equation}\label{eq:EBtest}
\Phi  \; {\rm is\ EB} \iff  (\Phi \otimes I) |\Psi \rangle_{sa} \langle \Psi|  \; {\rm is\ separable}\,,
\end{equation}
where $I$ is the identity map on an arbitrary ancillary system $a$, and $|\Psi \rangle_{sa}$ is a maximally entangled state of the bipartite system composed by the considered system $s$ and $a$. 

\section{Experimental Implementation}
Our experimental scheme adopts a Sagnac interferometric source of polarization-entangled photons \cite{source} in a Bell state $\ket{\Psi}_{sa}=\frac{1}{\sqrt{2}}\left(\ket{0}_{s}\ket{1}_{a}+e^{i\phi}\ket{1}_{s}\ket{0}_{a}\right)$, $\ket{0}\equiv\ket{H}$ ($\ket{1}\equiv\ket{V}$) being  the horizontal (vertical) polarization. The photons are generated in two indistinguishable Type-II \textit{parametric down-conversion} processes inside a PPKTP nonlinear crystal. Here, the photons belonging to a continuous wave laser of $405nm$ are converted into twin photons of $810nm$ at a rate $>60000 \, pairs/s$ and heralded efficiency $>16\%$ revealed by two synchronized \textit{avalanche photo-detectors} (APDs). The high purity of the generated state was measured, giving a fidelity of $F_{exp}=0.980\pm0.016$ to $\ket{\Psi}_{sa}$ \cite{entanglement} and a concurrence of $C_{exp}=0.973\pm0.004$ \cite{concurrence}. 
In order to take into account the non perfect purity of the input entangled state, we can model it as a Werner state $\rho_{W}=\frac{4F-1}{3}\rho_{sa}+\frac{1-F}{3}\mathbb{I}_{s}\otimes\mathbb{I}_{a}$ \cite{wernerstate,channelcapacity}, where $\mathbb{I}_x$ is the identity operator associated to subsystem $x$.

\begin{figure}[h]
	\centering
		\includegraphics[width=0.45\textwidth]{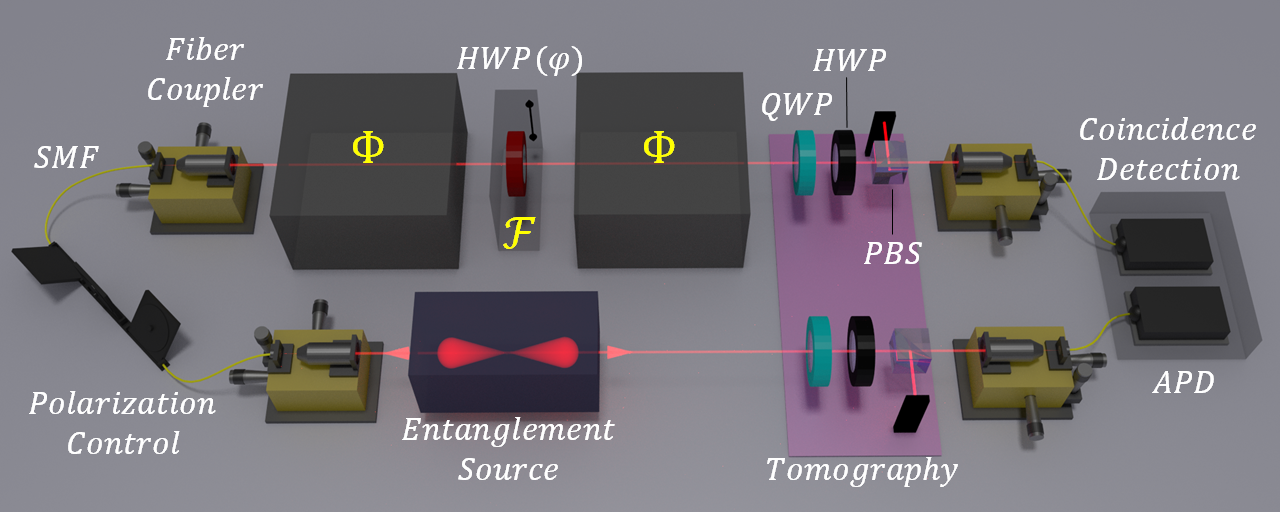}
	\caption{\textbf{Scheme of the experimental setup} A source of polarization entangled qubits sends the $a$-photon directly to the tomography stage, while the $s$-photon is transmitted by a SMF to the simulation of $\Omega=\Phi \circ \Phi$ or $\Omega'=\Phi \circ {\cal F} \circ\Phi$ and then measured in the same temporally synchronized bipartite tomography. $\Phi=\Phi_{PD}$ or $\Phi=\Phi_{AD}$ are represented by the black boxes, while $\mathcal{F}=\Lambda_\varphi$ is represented by the transparent-gray box enclosing a half wave plate $HWP(\varphi)$.  Finally, we specify that $\Omega$ ($\Omega'$)  corresponds to  the absence (presence) of $HWP(\varphi)$.}
	\label{setup}
\end{figure}

In order to test condition \eqref{eq:EBtest} in the laboratory, the $s$-photon (i.e. the photon which embodies the system $s$) is injected in a bulk optics setup that implements $\Omega=\Phi\circ\Phi$ or $\Omega'=\Phi\circ\mathcal{F}\circ\Phi$, while the $a$-photon (i.e. the ancillary system) propagates in free space. The bipartite output state $\rho_{sa}^{out}$ is measured by a hyper-complete tomography setup~\cite{tomography} (see Fig. \ref{setup}). Then,  the  $s$- and $a$-photons are coupled into \textit{single-mode fibers} (SMF) directly connected to an APD. The setup allows to measure the degree of entanglement remaining after the action of each implemented map.

\begin{figure}[h]
	\centering
	a) \includegraphics[width=0.195\textwidth]{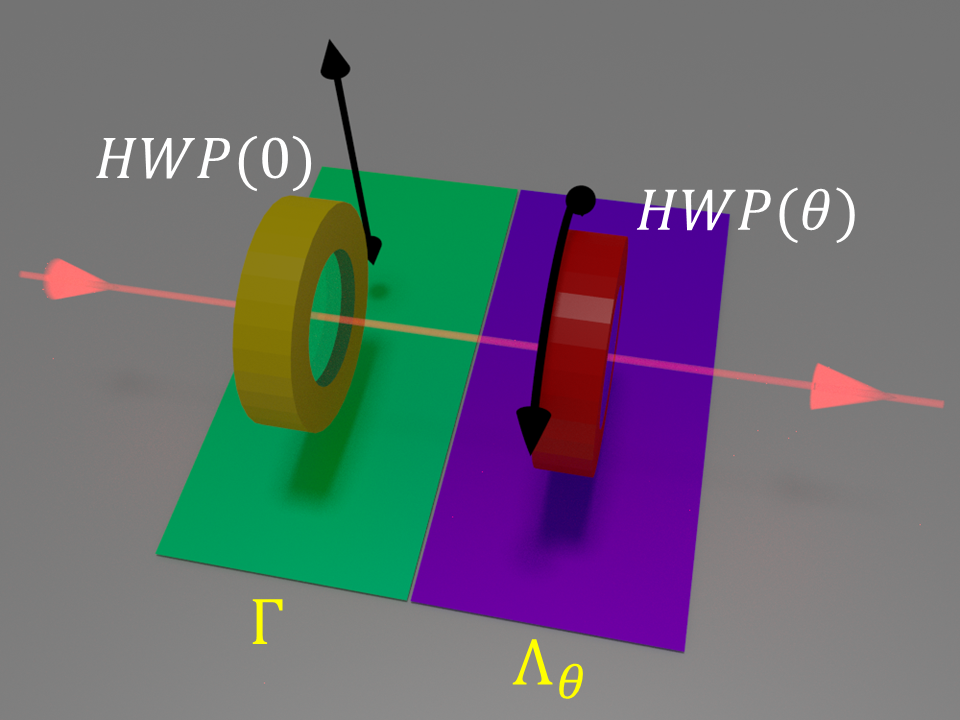}
	b) \includegraphics[width=0.22\textwidth]{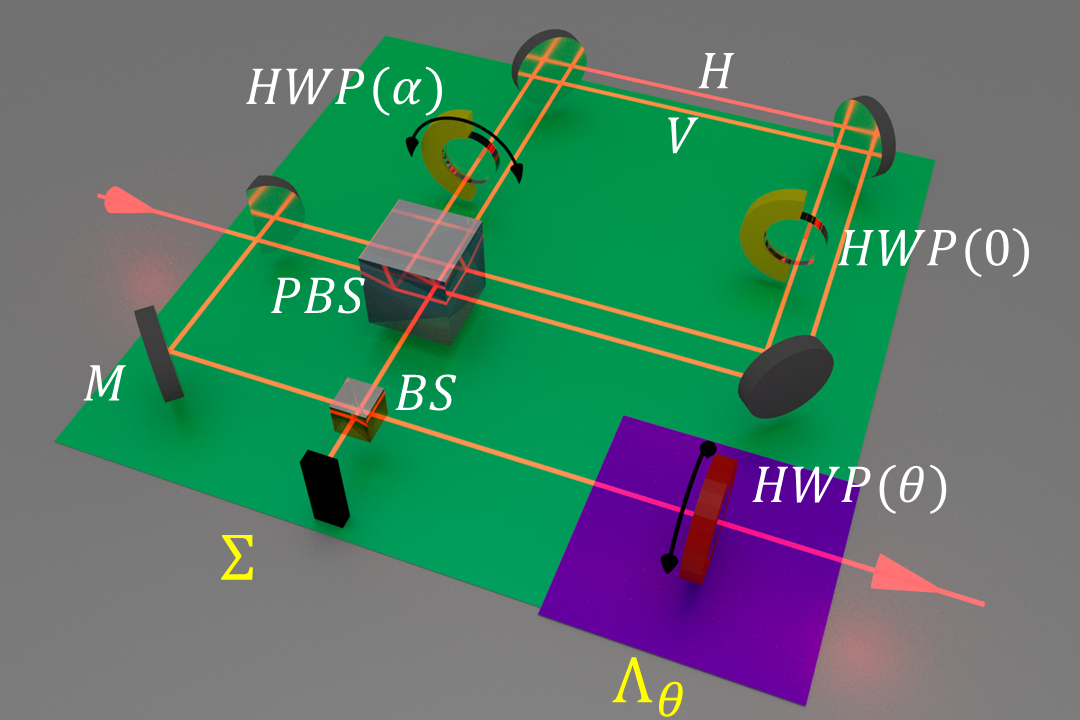}
	\caption{\textbf{Single channel modules.} a) $\Phi_{PD}$:  The unrotated yellow plate $HWP(0)$ constitutes the PD channel $\Gamma$, since $\mathbb{I}$ is applied when it is absent or $\sigma_{z}$ when it is present. b) $\Phi_{AD}$: The SI and MZI constitute the AD channel $\Sigma$, since transform the vertical polarization into horizontal by a rotation of $HWP(\alpha)$. Either in a) or b), the rotating red plate $HWP(\theta)$ represents $\Lambda_{\theta}$.}
	\label{channels}
\end{figure}

\textbf{Rotating Phase Damping Channel}: To simulate each $\Gamma$ channel (Eq.~\eqref{PDmap}), two operations are needed, $\mathbb{I}$ and $\sigma_{z}$.  They are simply implemented by the absence or presence of a \textit{half-wave plate} (HWP) fixed at $0$ degrees in the optical path of the photon, respectively. To simulate $\Lambda_\theta$ another HWP is permanently placed after $\Gamma$, but with a rotation degree of freedom in the angle  $\theta$  (as seen in Fig. \ref{channels} a)) \cite{channelcapacity,pauli_channel}.

Since both $\Lambda_\theta$ plates are synchronized in their rotation angle, there are only four combinations of Pauli operations; when the first $\Phi_{PD}$ is applying $\mathbb{I}$, the second $\Phi_{PD}$ can apply $\mathbb{I}$ or $\sigma_z$; when the first $\Phi_{PD}$ is applying $\sigma_z$, the second $\Phi_{PD}$ can apply $\mathbb{I}$ or $\sigma_z$. Then, the statistical mixture between $\mathbb{I}$ and $\sigma_{z}$ is obtained by extracting a fraction $P_{\mathbb{I}\mathbb{I}}=(1-\frac{p}{2})^{2}$ of coincidences from the $\mathbb{I}+\mathbb{I}$ tomography, a fraction $P_{\mathbb{I}\sigma_{z}}=(1-\frac{p}{2})\cdot\frac{p}{2}$ of coincidences from the $\mathbb{I}+\sigma_{z}$ and $\sigma_{z}+\mathbb{I}$ tomographies, and a fraction $P_{\sigma_{z}\sigma_{z}}=(\frac{p}{2})^{2}$ from the $\sigma_{z}+\sigma_{z}$ tomography.

Once the tomography registry fractions are combined, the new registry will be equivalent to a tomography of the state under a the action of $\Omega_{PD}=\Phi_{PD} \circ \Phi_{PD}$.

\textbf{Rotating Amplitude Damping Channel}: To simulate each $\Sigma$ channel, we use a displaced \textit{Sagnac interferometer} (SI), opened and closed by a single \textit{polarizing beam splitter} (PBS) (as seen in Fig. \ref{channels} b)). The parallel trajectories of $\ket{V}$ and $\ket{H}$ projections inside the SI are temporally compensated and go in clockwise and counter-clockwise directions, respectively. Both trajectories are intercepted by independent HWPs, a rotating one $HWP(\alpha)$ for $\ket{V}$ and another unrotated $HWP(0)$ for $\ket{H}$. The rotation angle $\alpha$ is related to the damping parameter $\eta$ by the expression $\alpha(\eta)=\frac{arccos(-\sqrt{1-\eta})}{2}$ \cite{cutandpaste,channelcapacity}.

After the SI there is an unbalanced \textit{Mach-Zehnder interferometer} (MZI), that allows to couple in the same trajectory the damped and undamped polarizations as they pass through a \textit{beam splitter} (BS). The temporal difference between the MZI arms is set to a value larger than the coherence length of the photons in order to simulate random phase fluctuations that destroy quantum interferences at its output.
 
The action of $\Omega_{AD}=\Phi_{AD} \circ \Phi_{AD}$ is then obtained by selecting the same damping $\eta$ in both $\Sigma$, while both HWPs corresponding to $\Lambda_{\varphi}$ rotate in a synchronous way.

\textbf{Filtering}: The protocol firstly requires to fix the damping parameter $p$ for $\Omega_{PD}$ or $\eta$ for $\Omega_{AD}$, to scan the channel in the rotation angle $\theta$ and verify the location of periodic EB regions.  It results that these regions are located around $\theta_{PD}=\frac{\pi}{8}\pm n\frac{\pi}{4}$ for the $(\Omega_{PD,s}\otimes\mathbb{I}_{a})(\rho_{sa})$ experiment and around $\theta_{AP}=\frac{\pi}{4}\pm n\frac{\pi}{2}$ for the $(\Omega_{AD,s}\otimes\mathbb{I}_{a})(\rho_{sa})$ experiment, in both cases with $n\in \mathbb{N}$. Once this condition is experimentally certified, one proceeds to fix the angle $\theta=\theta_{PD}$ or $\theta=\theta_{AD}$. Then, the operation of $\mathcal{F}$ is studied by scanning the rotation $\varphi$ of an extra HWP (as seen in Fig. \ref{setup}). 

As a consequence, either $\Omega_{PD}'  =\Phi_{PD} \circ \Lambda_{\varphi} \circ \Phi_{PD}$ or $\Omega_{AD}' =\Phi_{AD} \circ \Lambda_{\varphi} \circ \Phi_{AD}$ will be no more $EB$, in a region where, on the contrary, $\Omega_{PD}$ and $\Omega_{AD}$ were EB.

\section{Results}

\begin{figure}[h!]
	\centering
	a)	\includegraphics[width=0.45\textwidth]{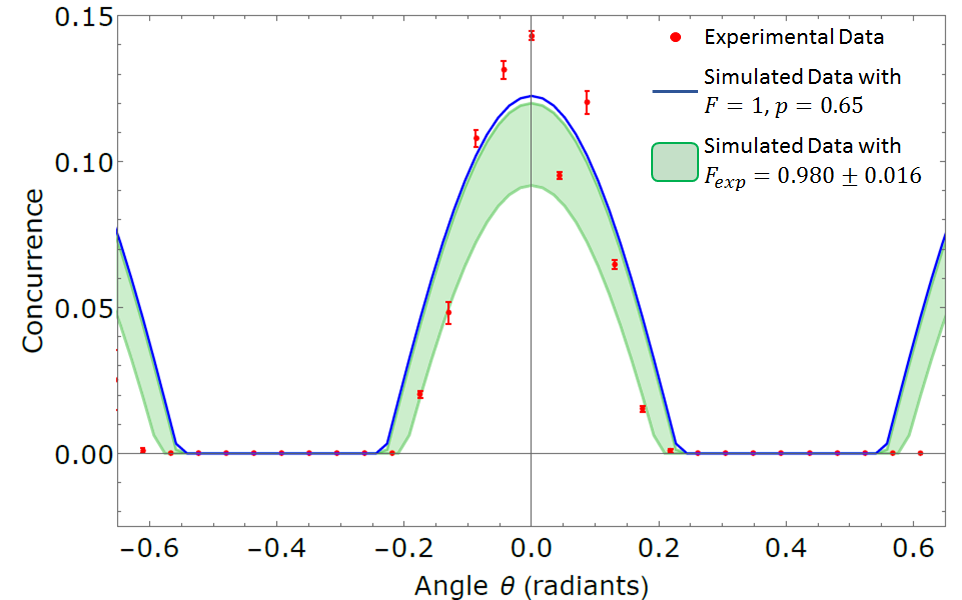}\\
	b)	\includegraphics[width=0.45\textwidth]{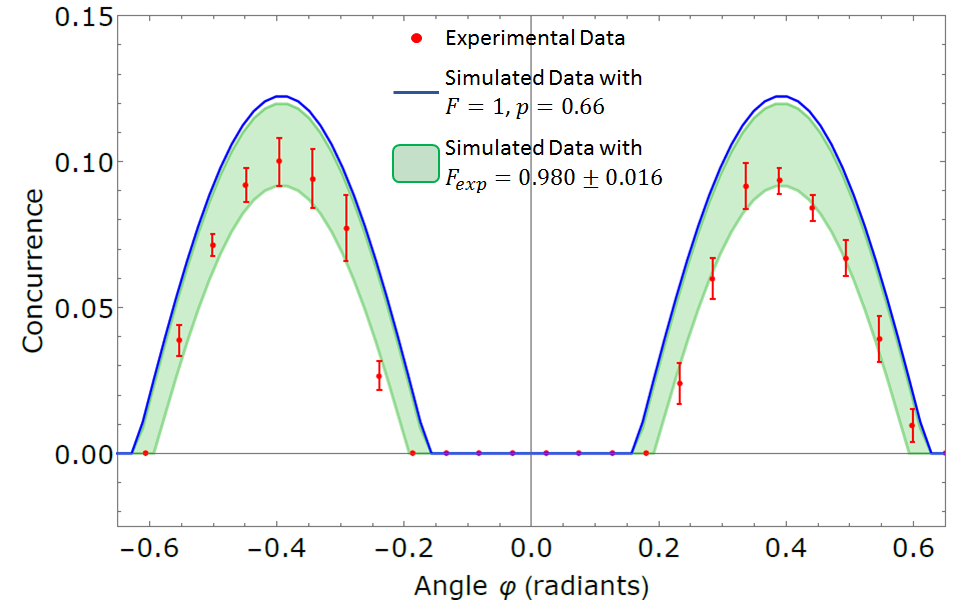}
	\caption{\textbf{Concurrence vs. PD mapping for $p=0.65$.} Red points represent the experimental data. Blue lines represent the simulated data for \textit{perfect optical elements} (POE) and pure entangled state with $F=1$. Green shaded areas represent the regions of simulated data for \textit{realistic optical elements} (ROE) and a mixed entangled state within one standard deviation of the fidelity $F_{exp}=0.980\pm0.016$. \textbf{a) $\Omega_{PD}$:} obtained by rotating $\theta$, with \textbf{EB behavior around $\theta=\pm\frac{\pi}{8}$}. \textbf{b) $\Omega'_{PD}$:} obtained by rotating $\varphi$, with a \textbf{revival of entanglement around $\varphi=\pm\frac{\pi}{8}$, while $\theta$ is fixed at $\frac{\pi}{8}$.}}
	\label{results_dephasing}
\end{figure}

\begin{figure}[h!]
	\centering
	a)	\includegraphics[width=0.45\textwidth]{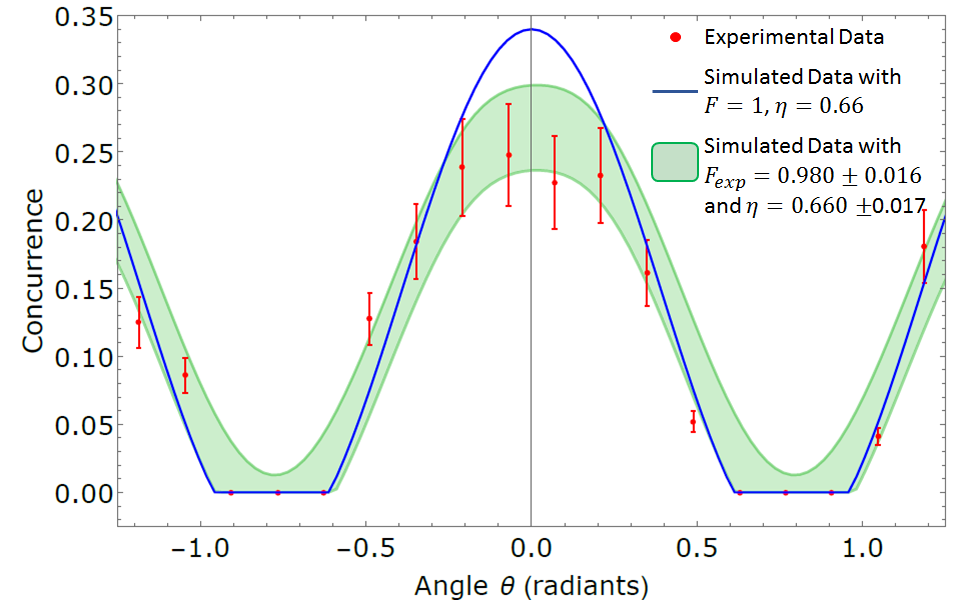}\\
	b)	\includegraphics[width=0.45\textwidth]{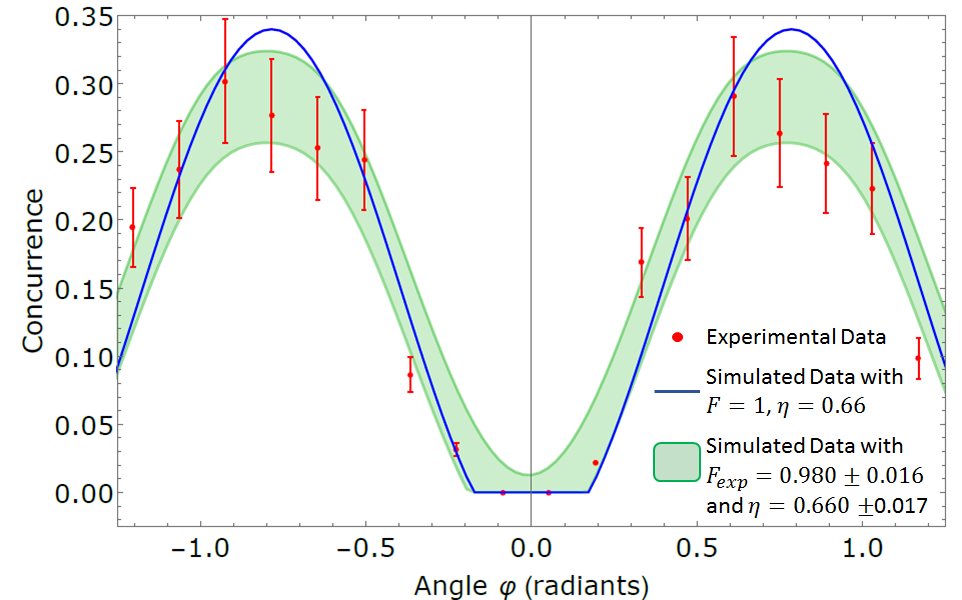}
	\caption{\textbf{Concurrence vs. AD mapping for $\eta=0.66$.} Red points represent the experimental data. Blue lines represent the simulated data for perfect optical elements (POE) and pure entangled state with $F=1$. Green shaded areas represent the regions of simulated data for realistic optical elements (ROE) and a mixed entangled state within one standard deviation of the fidelity $F_{exp}=0.980\pm0.016$ and the propagated error of the damping $\eta=0.66\pm0.017$. \textbf{a) $\Omega_{AD}$:} obtained by rotating $\theta$, with \textbf{EB behavior around $\theta=\pm\frac{\pi}{4}$}. \textbf{b) $\Omega'_{AD}$:} obtained by rotating $\varphi$, with a \textbf{revival of entanglement around $\varphi=\pm\frac{\pi}{4}$, while $\theta$ is fixed at $\frac{\pi}{4}$.}}
	\label{results_amplitude}
\end{figure}

In Fig.~\ref{results_dephasing} we report the experimental results  for the channels $\Omega_{PD}$ and $\Omega'_{PD}$ acting over the $s$-photon of  a pair of entangled photons, with the damping parameter set to the value $p=0.65$. Fig.~\ref{results_dephasing} a) shows the EB behaviour of $\Omega_{PD}$ around $\theta_{PD}=\pm\tfrac{\pi}{8}$ as predicted by the simulated model, while Fig.\ref{results_dephasing} b) shows an entanglement revival of $\Omega'_{PD}$ for $\varphi=\pm\tfrac{\pi}{8}$.
Similarly, in Fig.~\ref{results_amplitude} we report the results of $\Omega_{AD}$ and $\Omega'_{AD}$ channels acting over the $s$-photon from a pair of entangled photons, having set the damping parameter to the value $\eta=0.66\pm0.017$. Fig.~\ref{results_amplitude} a) shows the $EB$ behavior of $\Omega_{AD}$ around $\theta_{AD}=\pm\tfrac{\pi}{8}$ as predicted by the simulated model, while Fig.\ref{results_amplitude} b) shows an entanglement revival of $\Omega'_{AD}$ for $\varphi=\tfrac{\pi}{8}$. 

The experimental data were obtained by averaging and calculating the standard deviation over 5 values per point. The blue lines were calculated considering perfect input state and optical conditions. The simulated shaded green areas correspond to the regions of all possible experimental results within one standard deviation of $F_{exp}=0.980\pm0.016$ for $\Omega_{PD}$ and $\Omega'_{PD}$, and also considering the error propagation of 0.5 degrees of uncertainty in $\theta$ for $\Omega_{AD}$ and $\Omega'_{AD}$. This difference in the data analysis between PD and AD channel origins from the negligible error contribution of $0.5$ degrees of uncertainty in $PD$ channels. All $C_{exp}=0$ values have error bars within the size of the point.

The simulated data considered two scenarios: one with \textit{perfect optical elements} (POE) and maximally pure entangled input state, and another with \textit{realistic optical elements} (ROE), non maximally pure entangled input state and error propagation. The differences between these two cases are described in TABLE \ref{table}.

\begin{table}[h!]
\centering
		\begin{tabular}{|c|c|c|}
		\hline
			&POE & Average ROE\\
			\hline
			Fidelity&1&$0.980\pm0.016$\\
			\hline
			$TH_{BS}$&0.5&$0.507\pm0.016$\\
			$RH_{BS}$&0.5&$0.407\pm0.011$\\
			$TV_{BS}$&0.5&$0.495\pm0.018$\\
			$RV_{BS}$&0.5&$0.410\pm0.001$\\
			$TH_{PBS}$&1&$0.965\pm0.001$\\
			$RH_{PBS}$&0&$0.008\pm0.004$\\
			$TV_{PBS}$&0&$0.024\pm0.014$\\
			$RV_{PBS}$&1&$0.928\pm0.035$\\
			\hline
		\end{tabular}
		\caption{\textbf{Optical parameters and state fidelities.} The acronym POE (or ROE) is for perfect (or real) optical elements. TH and TV represent the the transmissivities in the horizontal and vertical polarizations respectively. RH and RV represent the the reflectivities in the horizontal and vertical polarizations respectively.}
		\label{table}
\end{table}

In both $PD$ (Fig.\ref{results_dephasing}) and $AD$ (Fig.\ref{results_amplitude}) cases there is a good agreement between experimental and simulated data. The discrepancies existing between $\Omega_{PD}$, $\Omega'_{PD}$ and their simulations could be attributed to the post-processing generation of the channel, since their action has been simulated by combining $\mathbb{I}$ and $\sigma_z$ operations with unstable photon counts during long-time scans varying $\theta$ and $\varphi$ rotation angles. On the other hand, discrepancies between $\Omega_{AD}$, $\Omega'_{AD}$ and their simulations are strongly related to the difficulty of coupling the 16 possible spatial modes within a unique SMF at the end of the entire channel.

\section{Conclusions}

We have given an experimental proof for a method  aiming at increasing the entanglement transmission distance, applicable to communication lines decomposable into elementary maps. 
The technique consists in placing appropriate unitary operations between the elementary steps of the channel, allowing to restore the entanglement transmissivity of an initially entanglement-breaking communication line.

We implemented different single-qubit channels acting on the polarization of single photons combining standard bulk optics elements: rotated phase plates, beam splitters and photo-detectors. We applied such channels to a subsystem of a maximally entangled state, and computed the remaining fraction of entanglement after performing a two-photon state tomography. We measured clear revivals of the output entanglement whenever a unitary filter, consisting in a phase-plate with appropriate rotation angle, was placed in the middle of the transmission process. 

Our results could be extended to more general physical scenarios involving different single qubit operations \cite{depasquale2012}, continuous and non-unitary channels \cite{cutandpaste}, or amendable Gaussian maps \cite{depasquale2013}.

\section{Acknowledgements}
\'A. Cuevas would like to thank the support from the Chilean agency CONICYT and to its PhD scholarships program.

J. Ferraz would like to thank the support from the Brazilian agency CAPES and the CsF program.

\end{document}